\documentclass[final,3p,times,twocolumn]{elsarticle}

\usepackage{graphicx}
\usepackage{amssymb}
\biboptions{comma,square}

\journal{High Energy Density Physics}

\begin{document}

\begin{frontmatter}



\title{Effects of Strong Magnetic Fields on Photoionised Clouds}


\author[Bonn,Dias]{Jonathan Mackey}
\author[Dias]{Andrew J.\ Lim}

\address[Bonn]{Argelander-Institut f\"ur Astronomie, Auf dem H\"ugel 71, 53121 Bonn, Germany.}
\address[Dias]{Dublin Institute for Advanced Studies, 31 Fitzwilliam Place, Dublin 2, Ireland.}

\begin{abstract}
Simulations are presented of the photoionisation of three dense gas clouds threaded by magnetic fields, showing the dynamical effects of different initial magnetic field orientations and strengths.
For moderate magnetic field strengths the initial radiation-driven implosion phase is not strongly affected by the field geometry, and the photoevaporation flows are also similar.
Over longer timescales, the simulation with an initial field parallel to the radiation propagation direction (parallel field) remains basically axisymmetric, whereas in the simulation with a perpendicular initial field the pillar of neutral gas fragments in a direction aligned with the magnetic field.
For stronger initial magnetic fields, the dynamics in all gas phases are affected at all evolutionary times.
In a simulation with a strong initially perpendicular field, photoevaporated gas forms filaments of dense ionised gas as it flows away from the ionisation front along field lines.
These filaments are potentially a useful diagnostic of magnetic field strengths in H~\textsc{ii} regions because they are very bright in recombination line emission.
In the strong parallel field simulation the ionised gas is constrained to flow back towards the radiation source, shielding the dense clouds and weakening the ionisation front, eventually transforming it to a recombination front.
\end{abstract}


\end{frontmatter}

% \linenumbers

\section{Introduction}
\label{intro}
Massive stars are born in cold and dense molecular clouds with gas temperatures of $T\approx15-50$ K, but very early in their lives such stars reach surface effective temperatures of $T_{\mathrm{eff}}\approx30\,000-50\,000\,$K and emit ionising photons at rates of $10^{47}-10^{50}$ s$^{-1}$ \cite{DiaFraSho98}, rapidly ionising their environment.
For Galactic heavy element abundances, the balance of photo-heating and radiative cooling gives $T\approx5\,000-10\,000$ K in photoionised gas \cite{Spi54}.
This heated region is hence a very over-pressurised  bubble of ionised gas around newly formed massive stars
(the ratio of ionised gas pressure to that in surrounding neutral gas $\approx 10^2-10^3$),
and is known as an H~\textsc{ii} region.
The bubble expands by driving a shock outwards \cite{Spi78}; when this shock is isothermal a thin, dense, and generally unstable shell forms at the H~\textsc{ii} region boundary.

The interstellar medium (ISM) in molecular clouds is generally observed to be clumpy and filamentary \cite{HilMotDidEA12,SchCseHenEA12} so, even in the absence of instabilities, the expansion rate of the H~\textsc{ii} region will be uneven and angle-dependent.
Indeed, some of the most striking and easily observed optical nebulae are produced by recombination radiation from ionisation fronts (I-fronts) around pillars of neutral gas, also known as elephant trunks \cite{HesScoSanEA96}.
Younger H~\textsc{ii} regions appear quite regular and spherical, e.g.\ RCW 120 \cite{DehZavSchEA09}, whereas more mature H~\textsc{ii} regions with ages $\tau\geq1-2$ Myr such as the Eagle Nebula \cite{HesScoSanEA96,GraPou12}, IC 1396 \cite{BarVinDreEA11}, and the Carina Nebula \cite{Har12} contain well-developed pillars and globules.
It has been unclear for decades whether ionisation front instabilities or pre-existing environmental inhomogeneities are primarily responsible for forming these pillars and globules \cite[e.g.][]{Kah58,WilWarWhi01}; both processes should be acting, but possibly on different length scales.

Magnetic fields have been suggested as the driving process for apparently helical  and rotating structure in some elephant trunks \cite{GahCarJohEA06}, suggesting that magnetic fields could be dynamically important in these structures.
Only one measurement of the field orientation in pillars has been made so far, however, for the Eagle Nebula by \cite{SugWatTamEA07}.
Cometary globules are related structures; magnetic field measurements in some of these \cite{MarFor78,SriBhaRaj96,BhaMahMan04} show similar field alignment with the head-tail morphology of the globules (but see \cite{Bha99} for a counterexample).

A magnetic field provides additional pressure support to the gas (which cannot be lost by radiative cooling, unlike thermal pressure), and it constrains gas to flow along field lines (for both neutrals and ions in the ideal magnetohydrodynamic [MHD] limit, because of efficient collisional coupling).
Pioneering calculations by \cite{Las66b} suggested their effects would be significant in H~\textsc{ii} regions; analytic models of photoionised globules including an approximate magnetic pressure \cite{BerMcK90} showed that it can be important to the pressure support of such clouds.
Extra magnetic pressure support for the Eagle Nebula pillars was also deduced by \cite{RyuKanMizEA04} by analysis of the gas densities and temperatures inferred from observations.
Magnetostatic turbulence \cite{RyuRem07} was proposed as a possible source of long-term pressure support, as supersonic turbulence decays very rapidly.
Alternatively, the pillars may not be in pressure equilibrium but could still be dynamically evolving structures \cite{WhiNelHolEA99,MacLim10}.

The structure of magnetised 1D I-fronts was studied by \cite{RedWilDysEA98}, and more recently \cite{Wil07} used 2D ionising radiation-MHD (IR-MHD) simulations to study slab-symmetric I-fronts with various initial magnetic field configurations, finding that strong magnetic fields dramatically changed the evolution of an advancing I-front.
The photoionisation of a dense spherical clump in a magnetised medium was simulated in 3D by \cite{HenArtDeCEA09}.
They found that weakly magnetised clumps with initial field strength $B=59$ micro-Gauss ($\mu$G) evolved similarly to the purely hydrodynamic case, whereas strongly magnetised clumps ($B=186\ \mu$G) evolved in a very different manner.
The compression and structure of the neutral globule changed and the flow of ionised gas from the I-front was highly constrained.
For a model with a magnetic field perpendicular to the radiation propagation direction, the ionised gas was confined to a dense ribbon or filament standing off from the globule, which eventually shielded the globule from much of the ionising radiation.
This work was followed up by \cite{MacLim11}, who used a similar model to \cite{HenArtDeCEA09} to study the photoionisation of multiple clumps in different configurations.
They confirmed most of the results of \cite{HenArtDeCEA09}, and the additional asymmetry introduced by the multiple clumps led to pillar fragmentation in one case.
It was also shown in \cite{MacLim11} that the dense filament of ionised gas could be explained quite simply by analysis of the jump conditions for isothermal magnetised shocks.
Both studies found that, for a clump with a weak magnetic field oriented perpendicular to the radiation propagation direction initially (hereafter referred to as perpendicular field models), the magnetic field was swept into alignment with the pillar/globule during its evolution.

In this paper two photoionisation simulations with magnetic fields initially aligned with the radiation propagation direction (hereafter parallel field models) are presented and compared to the perpendicular field simulations already described in \cite{MacLim11}, for medium and strong initial magnetic fields.
In the next section the numerical methods and simulation setup will be reviewed briefly; the results of the simulations will be presented in section ~\ref{sec:results}; the results will be discussed in the context of observations and previous work in section~\ref{sec:discussion}; and conclusions will be presented in section~\ref{sec:conclusions}.

\section{Methods}
\label{sec:methods}
\subsection{Code Description}

The IR-MHD code and simulation setup are described in detail in \cite{MacLim10,MacLim11}, including results of tests of the algorithms, so only the essential features of the code are reviewed here.
The ideal MHD equations (using an ideal gas equation of state with $\gamma=5/3$) are solved on a uniform 3D grid using the second-order-accurate (in time and space), finite-volume scheme of \cite{FalKomJoa98}, with inter-cell fluxes calculated using the conserved variables Riemann solver of \cite{CarGal97} (implemented as in \cite{StoGarTeuEA08}).
The mixed-GLM method \cite{DedKemKroEA02} is used to control errors from non-zero divergence of the magnetic field.

Microphysical processes that modify the effective equation of state of the gas (radiative heating and cooling, ionisation and recombination) are included by operator-splitting; in the source-term update the thermal pressure ($p_g$) and hydrogen ion fraction ($y$) are integrated forwards in time on a point-local cell-by-cell basis.
Photoionisation is calculated using the `on the spot' approximation, where recombinations to the ground state of H are ignored by assuming they result in re-emission of an ionising photon which is absorbed locally.
This allows one to ignore scattered radiation and calculate photoionisation rates based only on the direct photon flux from point sources (see \cite{WilHen09} for a recent discussion of this approximation).
Attenuation of photons is calculated using a short-characteristics raytracer \cite{MelIliAlvEA06}.

This algorithm is not radiation-MHD in the sense that radiation is not included as part of the total energy density, and radiation pressure is not considered.
Rather it is a photon counting scheme that tracks where ionising photons are absorbed and deposit their energy.
This photoionisation heating is the dominant effect of ionising radiation for the situations we are considering.

\subsection{Simulation Setup} 
The essential properties of the suite of simulations are stated below, in particular the radiation source, the computational grid, and the initial conditions for gas and magnetic field.
The reader is referred to \cite{MacLim11} for further details.
The simulations to be described here are models R5, R6, R8, and R9 from table~2 in \cite{MacLim11}; their different initial magnetic field configurations are given in Table~\ref{tab:rmhd_Fields}.
For all simulations, an ionising photon source with luminosity $Q_0=2\times10^{50}$ photons per second is placed at the origin; the simulation domain is $x\in[1.5,6.0]$ parsecs (pc, $3.086\times10^{18}\,$cm), and $(y,z) \in [-1.5,1.5]$ pc resolved with $384\times256^2$ grid cells.
Three clumps of gas with mass $M\approx28\,\mathrm{M}_{\odot}$ are placed at positions (in pc) $(2.3,\,0,\,0)$, $(2.75,\,0,\,0.12)$, and $(3.2,\,0,\,-0.12)$, with peak overdensities of 500 compared to the background ISM which has H number density $n_{\mathrm{H}}=200\,\mathrm{cm}^{-3}$.
The clumps have Gaussian profiles of width $\sigma=0.09$ pc, and the simulation is initially in pressure equilibrium and static.
The four simulations differ only in the initial magnetic field strength and orientation (see Table~\ref{tab:rmhd_Fields}).
All four models use zero-gradient boundary conditions.
Models R5 and R6 have medium field strengths, magnetically dominated in neutral gas, but the much hotter photoionised gas is thermal-pressure dominated; for R8 and R9 both phases are magnetically dominated.
R5 and R8 have perpendicular initial field configurations, whereas R6 and R9 have a parallel field.

\begin{table}
  \centering
  \begin{tabular}{ | l | c  c  c | c c |}
    \hline
    Model & $B_x$ & $B_y$ & $B_z$ & $\beta_n$ & $\beta_i$ \\
    \hline
    R5 & 0 & 0 & 53.2 & 0.12 & 4.0 \\
    R6 & 53.2 & 0 & 0 & 0.12 & 4.0 \\
    R8 & 14.2 & 7.1 & 158.9 & 0.014 & 0.43 \\
    R9 & 159.5 & 7.1 & 10.6 & 0.014 & 0.43 \\
    \hline
  \end{tabular}
  \caption{Cloud and initial magnetic field configurations, boundary conditions
    applied, and final simulation times for the R-MHD simulations.
    Field strengths are quoted in micro-Gauss.
    Plasma parameter $\beta\equiv p_g/p_m$ (the ratio of thermal to magnetic pressure)
    is shown in columns 5 and 6: $\beta_n$ refers to the neutral gas initial conditions, and $\beta_i$ to photoionised gas at the background density and with $T=8000$ K.}
    \label{tab:rmhd_Fields}
\end{table}

\section{Results}
\label{sec:results}
\begin{figure*}
  \centering 
  \includegraphics[width=0.8\textwidth]{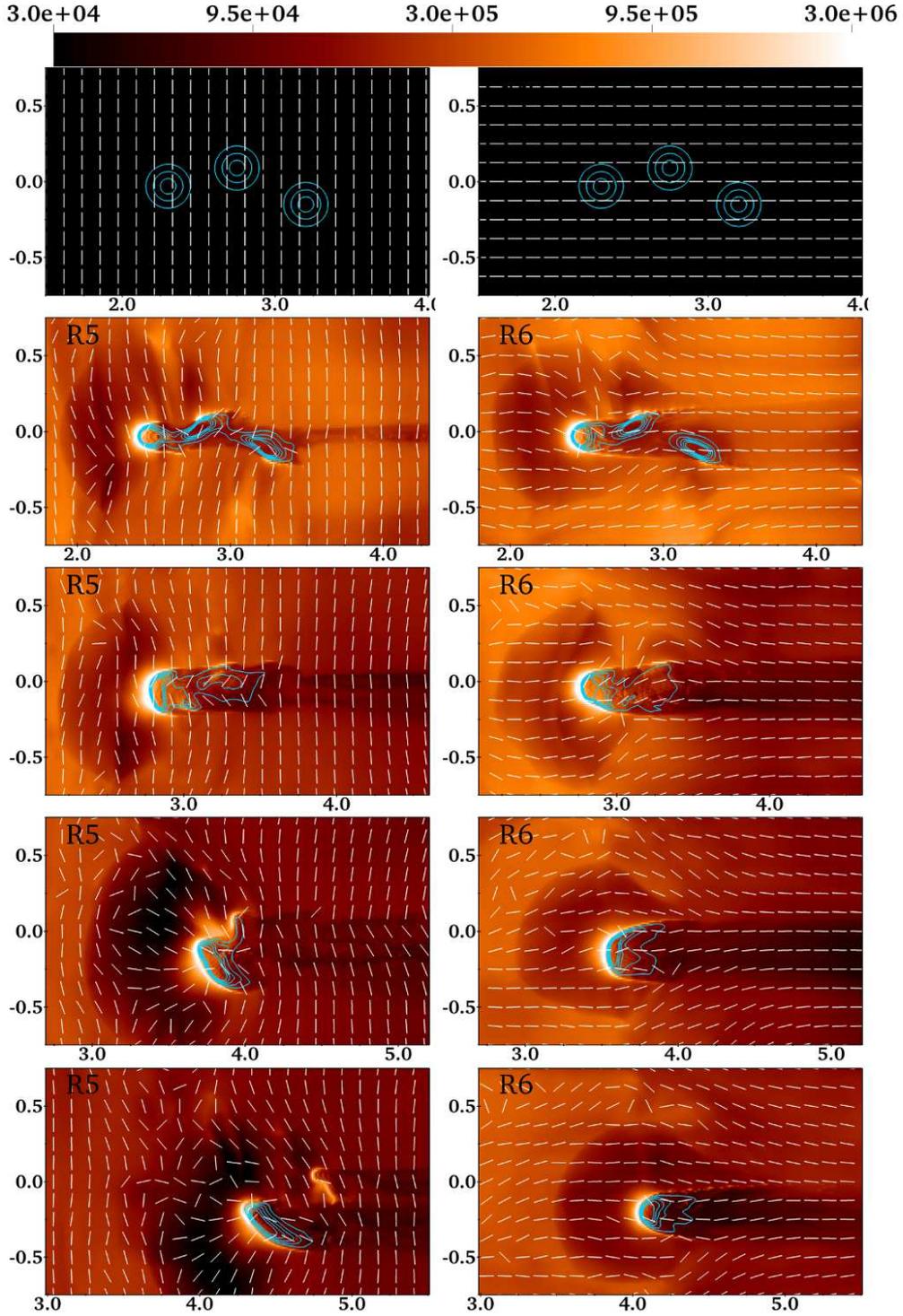}
  \caption{
    Projections through Simulations R5 (left) and R6 (right) at times (from top to bottom) 0, 100, 200, 400, 500 kyr.
    The image $x$- and $y$-axes are the simulation $x$- and $z$-axes; the line of sight is the simulation $y$-axis.
    The colour scale shows H$\alpha$ intensity, integrated along the line of sight assuming no background sources.
    White lines indicate projected magnetic field orientation at the midpoint of each line, and turquoise contours show projected neutral hydrogen density on a linear scale.  The five contours are at $N(H^0)=(0.2,0.4,0.6,0.8,1.0)\times 10^{21}$ cm$^{-2}$.
    Positions are shown on the axes labels in parsecs relative to the source; not all of the simulation is shown, and the image window moves further from the source from image to image.
    }
  \label{fig:R5R6}
\end{figure*}

\begin{figure*}
  \centering 
  \includegraphics[width=0.85\textwidth]{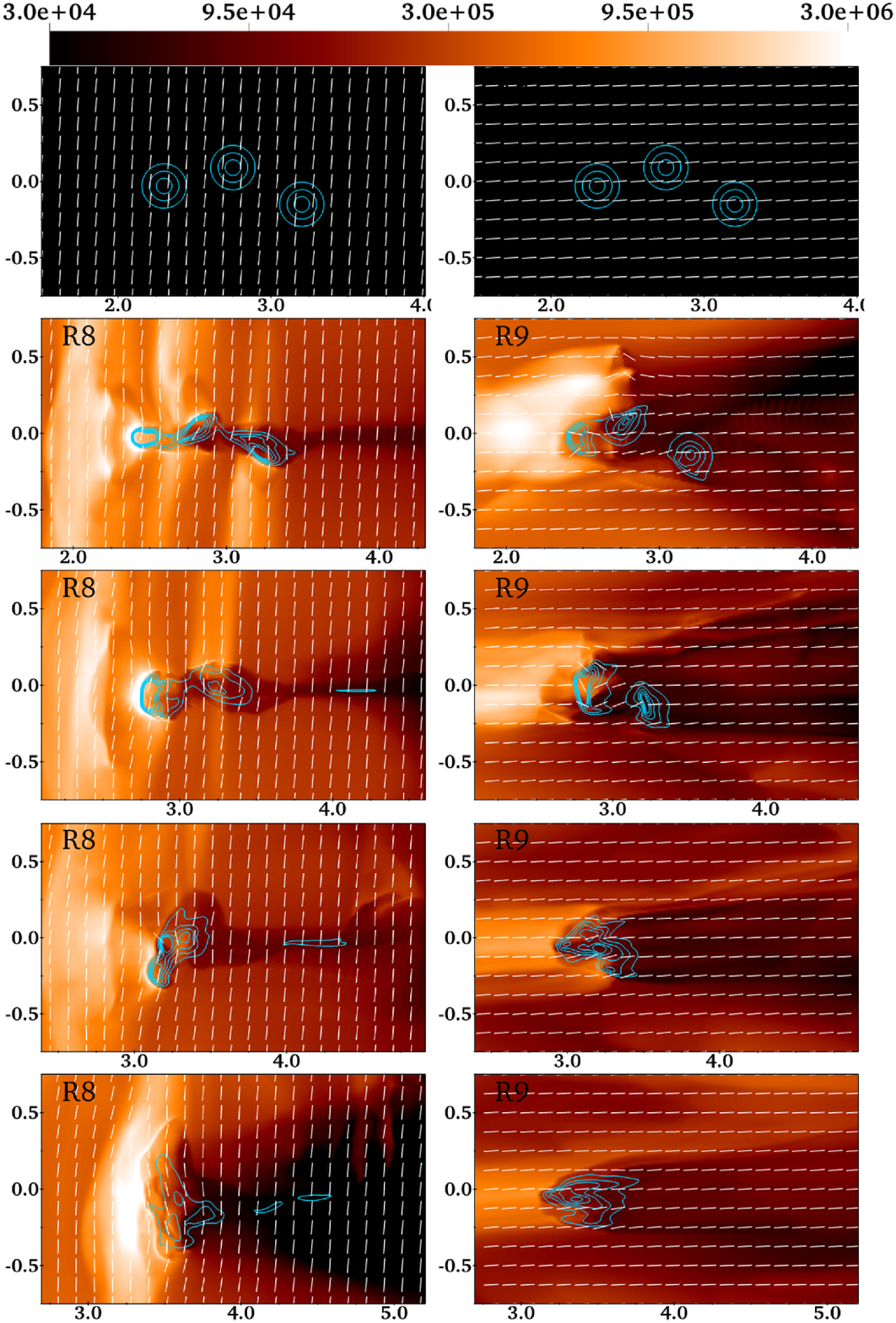}
  \caption{
    As Fig.~\ref{fig:R5R6} but showing projections through simulations R8 (left) and R9 (right) at times (from top to bottom) 0, 100, 200, 300, 400 kyr.
    The projection and the colour scales are the same as before.
    }
  \label{fig:R8R9}
\end{figure*}

Observable properties of the simulations have been calculated by projecting the 3D data onto a 2D plane, using weighted integrals along the line-of-sight (LOS).
Three observables are plotted in Figures \ref{fig:R5R6} and \ref{fig:R8R9} for a series of times: the intensity of the H$\alpha$ spectral line ($n=3\rightarrow2$ transition of H$^0$), with arbitrary normalisation; the neutral H column density, N(H$^0$), with units cm$^{-2}$, and the magnetic field orientation perpendicular to the line of sight.

Evaluation of N(H$^0$) is a simple summation of neutral gas along each column of grid cells.
The H$\alpha$ line intensity is calculated including dust attenuation but ignoring radiation scattered into the LOS~\cite{MacLim11}.  The gas temperature ($T\lesssim0.7$ eV) is too low to collisionally populate the energy levels, so all of the H$\alpha$ emission is from recombinations to excited states and the emissivity scales approximately with the recombination rate. 
The projected magnetic field is calculated using a Stokes parameter formalism, integrated along the LOS and transformed back to a magnetic field orientation~\cite{ArtHenMelEA11,MacLim11}; this mimics the polarimetry observations of \cite[e.g.][]{SugWatTamEA07}.

\subsection{Simulations R5 and R6 -- medium field strength}
Figure~\ref{fig:R5R6} shows projections along the simulation's $y$-axis, so the image plane is the simulation $x$-$z$ plane.
R5 is shown at left, with the initial perpendicular field fully in the image plane, and R6 is shown at right, with the initially parallel field again fully in the image plane.  The top two panels (at $t=0$) are dark because the gas is initially neutral; N(H$^0$) contours show the positions of the dense clumps.

The second row shows the situation at $t=100$ kyr, at the end of the radiation-driven implosion phase of evolution for the clump closest to the star.
The main features of the two simulations are similar at this stage, with a strong radial photoevaporation (ablation) flow from the main I-front visible as a rim of bright H$\alpha$ emission that weakens radially outwards because of the decreasing gas density.
These models have $\beta>1$ in ionised gas (where $\beta\equiv p_g/p_m$ is the ratio of thermal to magnetic pressure), so deviations in ionised gas flow as a consequence of the magnetic field are relatively minor.
The shadowed `cometary tail' behind the clumps is rather different, with strong compression into a sheet possible in R5, whereas the parallel field in R6 resists compression from all sides.
Neutral gas between the clumps is more strongly compressed in R5 than R6 for the same reason.

The two models remain similar at $t=200$ kyr (third row), but by $t=400$ kyr (fourth row) substantial differences have arisen.  
R6 is roughly axisymmetric but R5, with its strong $\hat{z}$ component of $\mathbf{B}$, allows vertical motion (in the image plane) much more readily than horizontal or line-of-sight motion.  
Neutral gas flows in R5, driven in response to the photoevaporation flows, are therefore no longer axisymmetric and the merged clump structure has fragmented into two cometary globules, one much larger than the other.

\subsection{Simulations R8 and R9 -- Strong field}
Simulations R8 and R9 are the same as R5 and R6 but with a $3\times$ stronger field, and small off-axis magnetic field components so that it is not perfectly grid aligned.
The same plots as in Figure~\ref{fig:R5R6} are shown for R8 and R9 in Figure~\ref{fig:R8R9}, now for times $t=0, 100, 200, 300$, and $400$ kyr.
It is immediately clear that the two models have completely different evolution from 100 to 400 kyr.

As discussed in \cite{HenArtDeCEA09,MacLim11}, the ambient magnetic pressure confines the photoevaporation flow to a much smaller volume for these field strengths, and after the termination shock the gas flow is approximately one-dimensional, along field lines.
For R8 (left panels) this leads to filaments of dense ionised gas moving toward the $\hat{z}$ boundaries.  
This dense layer shields the I-front at the clump surface, weakening the photoevaporation flow and reducing the intensity of H$\alpha$ emission at the I-front.
Indeed, the dense ribbons of ionised gas are the brightest regions in H$\alpha$ emission.
By 400 kyr the dense ionised layer allows significant recombination of the ISM at $x>4$ pc (the black region in H$\alpha$).
Neutral gas is also constrained to follow field lines and so the pillar structure is more anvil-shaped, elongated along the field direction.

In simulation R9, the photoevaporation flow pushes out spherically against the magnetic pressure to some extent, but beyond the termination shock the postshock gas then follows field lines back toward the radiation source.
This gas remains in a line between the radiation source and the dense clumps, accumulating until no more photons reach the dense clumps.  Even at 100 kyr (second row) significant recombination behind the ionised gas is clearly seen at $(x,z)\approx(2.75,0.3)$ pc, and the third clump at $(3.25,-0.12)$ remains well shielded and has barely been impacted by photoionisation.
A cometary tail has not formed to any extent because the shadowed region is basically incompressible (the tail in R8 at left is actually an edge-on thin sheet, because of the anisotropic magnetic pressure).
From 100-400 kyr the H$\alpha$ emission from the clump surface gets progressively weaker, and most of the emission is from the ionised `plug' of gas closer to the star.  Analysis of the gas flow at $t=400$ kyr shows that the clump boundary has actually become a recombination front, where ionised gas recombines, cools and gets denser, joining the clump material.
The gas velocity field has basically become one-dimensional over the whole simulation domain, very reminiscent of the 2D simulations of \cite[fig.~3]{Wil07}.

The early part of R9's evolution is qualitatively similar to the single-clump simulation S00L of \cite[fig.~8]{HenArtDeCEA09}, although the three clumps in our calculation make the solution much less symmetric.
Although it appears \cite{HenArtDeCEA09} did not follow the evolution to the point where the I-front in R9 switched to a recombination front, it is difficult to make a meaningful comparison based on simulation times because the parameters of the simulations are quite different.

\section{Discussion}
\label{sec:discussion}
It is important to ask whether or not analogues of these idealised models exist in reality, and also to what extent the numerical approximations made (ideal MHD, limited spatial resolution, and neglecting the diffuse ionising radiation field) are affecting the solution.
The first question can be addressed by more realistic simulations and detailed comparison to observations, and the second can potentially be addressed in laboratory experiments.
A review of the current status of high energy-density experiments with photoionised plasmas is given by \cite{Man12}; their application to modelling of pillar formation and I-front instabilities is discussed by \cite{KanMizPouEA05,KanSmaMarEA12}.
Extension of these experiments to include MHD effects requires an understanding of the magnetic properties of the materials used, but is in principle possible.
An important issue is that the limited numerical resolution of our calculations may artificially suppress any instabilities in the dynamics that are seeded on small scales (see \cite{MizKanPouEA06}), and laboratory experiments could certainly test this \cite{KanSmaMarEA12}.

In all cases where magnetic fields have been measured in interstellar clouds, the magnetic energy density is found to be comparable to kinetic and gravitational energy \cite[e.g.][]{MyeGoo88,MyeGooGusEA95,Cru99}, and larger than the thermal energy density.
Field strengths such as those in R5 and R6 are therefore the norm, but cases such as R8 and R9 may be somewhat extreme.
Simulations of photoionisation in a turbulent cloud \cite{ArtHenMelEA11} showed that situations such as R8/R9 did not arise naturally, with the caveat that the initial magnetic field strength in the simulation was set by the specific turbulence calculation they began with.

The high density of molecular pillars means that we are in the fully collisional MHD regime but the ion/electron fraction is typically very low, so imperfect ion-neutral coupling may be an important consideration on small scales.
Multi-fluid simulations of weakly ionised plasmas including non-ideal MHD effects \cite{JonDow12} are now beginning to test the limits of applicability of ideal MHD on small scales in molecular clouds.

Infrared observations are dramatically increasing our understanding of massive star formation regions \cite[e.g.][]{SchCseHenEA12,Har12}, and millimetre data can now probe gas properties with high sensitivity and spatial resolution \cite{GraPou12}.  Together with recombination-line data to study ionised gas, detailed studies of the interfaces between ionised and neutral gas can now be performed \cite{BohTapRotEA04,PelBalBroEA07}.  Examples of dense sheets or filaments of photoionised gas near the heads of pillars have been found in NGC 6357 \cite{BohTapRotEA04} and NGC 3603 \cite{BraGreChuEA00}.
Combined with velocity information in the ionised gas, and polarimetry data to constrain the magnetic field geometry, it would be possible to determine if these regions are magnetically dominated, and so to constrain the field strength.

\section{Conclusions}
\label{sec:conclusions}
Simulations have been presented of the photoionisation of three dense clumps threaded by magnetic fields of different field strengths and orientations.
These are very idealised models, designed to show the effects of different initial field orientations and strengths, and they confirm and extend previous 2D results \cite{Wil07}, and models of photoionisation of a single clump in 3D \cite{HenArtDeCEA09}.
For moderate initial magnetic field strengths ($\beta<1$ in neutral gas, $\beta>1$ in ionised gas) the initial radiation-driven implosion phase is not strongly affected by the field geometry, and the photoevaporation flows are also similar.
Over longer timescales, however, the parallel field model R6 remains basically axisymmetric whereas the pillar in the  perpendicular field model R5 fragments at late times in a direction aligned with the magnetic field.

For stronger initial fields ($\beta<1$ in both neutral and ionised gas), the gas dynamics in all gas phases are affected at all evolutionary times.
In the parallel field model (R9) the I-front of the dense cloud weakens and becomes a steady recombination front, because photoionised gas is constrained to remain in a line between the radiation source and the cloud (the gas flow becomes almost one-dimensional).
Filaments of dense ionised gas also develop in the perpendicular field model R8, as photoevaporated gas flows away from the I-front along field lines.
These filaments are potentially a useful diagnostic of magnetic field strengths in H~\textsc{ii} regions because they are so bright in recombination line emission.
The absence of this emission in the recombination front in model R9 may also provide useful observational constraints.



\section*{Acknowledgments}
JM acknowledges funding for this work from the Irish Research Council for Science, Engineering and Technology, and from the Alexander von Humboldt Foundation.
We thank the SFI/HEA Irish Centre for High-End Computing (ICHEC) for the provision of computational facilities and support. AJL thanks DIAS for the kind provision of a Research Associateship.

%% References


\bibliographystyle{elsarticle-num}
\bibliography{jm_hedla}






% \end{thebibliography}

\end{document}